\documentclass[12pt]{iopart}

\usepackage{url}
\usepackage{mathrsfs}
\usepackage{dcolumn,longtable}
\usepackage{graphicx}

\newcommand{\ket}[1]{\vert #1 \rangle}

\begin{document}
\title[EPR-enhanced crystal field analysis]{Electron paramagnetic resonance enhanced crystal field analysis for low point-group symmetry systems:\\ C$_{2v}$ sites in Sm$^{3+}$:CaF$_2$/SrF$_2$}

\author{S P Horvath$^{1,2}$, J-P R Wells$^{1,2}$, M F Reid$^{1,2}$, \\
   M Yamaga$^3$ and
   M Honda$^4$     
  }

\address{$^1$ School of Physical and Chemical Sciences, University of Canterbury, PB 4800, Christchurch 8140, New Zealand}
\address{$^2$ The Dodd-Walls Centre for Photonic and Quantum Technologies, New Zealand}

\address{$^3$ Department of Mathematical and Design Engineering, Gifu University, Gifu 501-1193, Japan}

\address{$^4$ Department of Science Education, Naruto University of Education, Naruto 772-8502, Japan}

\ead{sebastian.horvath@fysik.lth.se, jon-paul.wells@canterbury.ac.nz}

\date{\today}

\begin{abstract}
  We present a comprehensive spectroscopic study of  C$_{2v}$ point-group symmetry sites in Sm$^{3+}$:CaF$_2$/SrF$_2$ codoped with either NaF or LiF. Data includes  electron paramagnetic resonance measurements of Zeeman and hyperfine interactions for the ground state, as well as site-selective excitation and fluorescence spectroscopy up to the $^{4}${G}$_{5/2}$ multiplet. Inclusion of the EPR data allowed us to determine unique crystal-field parameters. The parameters provide information about the geometry of the sites and the nature of the interactions between the Sm$^{3+}$ dopant and the alkaline earth co-dopant. 
\\  \\
\today
\end{abstract}


\maketitle

\section{Introduction \label{sec:intro}}
Accurate crystal-field calculations of magnetic and hyperfine interactions for rare-earth ions in low point-group symmetry systems is a difficult problem  \cite{Guillot-NoelCalculationanalysishyperfine2010,sukhanov2018}.
While previous studies have employed ground-state $g$ values to enhance crystal-field calculations \cite{sukhanov2018,BakerEndor1970,antipin1972,freeth_zeeman_1982,falin2003}, accurately reproducing magnetic and hyperfine interactions for low point-group symmetry systems is still challenging. Crystal-field models for low-symmetry  materials such as yttrium orthosilicate have potential to aid in the development of the many recently demonstrated rare-earth ion based applications employing low-symmetry host materials. This includes advances for optical quantum memories, quantum-gate implementations, and optical-to-microwave modulators \cite{zhong_optically_2015,usmani2010,lovric2013,longdell_experimental_2004,rippe2008,probst2013,xavi2015}.

Here we present a detailed optical and magnetic characterization of the C$_{2v}$ centre in Sm$^{3+}$ doped crystals. 
The data is analysed  using a crystal-field fitting scheme that simultaneously employs optical fluorescence, magnetic Zeeman, and hyperfine data. Our parameters  are largely consistent with those for  Dy$^{3+}$ orthorhombic centres in alkaline-earth co-doped fluorides \cite{antipin1972}. This work demonstrates that  hyperfine interactions can be accurately modelled using a crystal-field approach in  low-symmetry systems.

There are two main substitutional sites for  Sm$^{3+}$ in CaF$_2$ and SrF$_2$: the tetragonal C$_{4v}$(F$^-$) centre and  cubic ($O_h$)  \cite{freeth_zeeman_1982,weber_paramagnetic_1964,ReevesSiteselectivelaserspectroscopy1992,wells_polarized_2000,MurdochSiteselectivespectroscopyTb1997,StricklandSiteselectivespectroscopyTm1997}. For crystals grown under oxidizing conditions, a C$_{3v}$(O$^{2-}$) centre is also present, having an O$^{2-}$ ion in a nearest neighbour position along the $[111]$ direction from the Sm$^{3+}$ ion. Co-doping of rare-earth doped alkaline earth fluoride crystals with Li$^+$ or Na$^+$ ions is known to enhance the concentration of cubic centres \cite{zakharch.bp_experimental_1966,jamison_sharp_1999} but also gives rise to rare earth centres charge compensated by monovalent alkali ions. Various configurations for these centres are given by Pack \emph{et al.} \cite{pack_ce3+:na+_1989} who studied the $4f^1 \to 4f^05d^1$ transitions of Na$^+$ codoped Ce$^{3+}$:CaF$_2$ and Ce$^{3+}$:SrF$_2$ crystals. While their work suggested the presence of multiple Na$^+$ charge compensated centres, a later study focusing on EPR intra- and inter- configurational transitions, of Ce$^{3+}$:CaF$_2$ optimally codoped with either Li$^+$ or Na$^+$, suggests the presence of three main centres \cite{yamaga_optical_2004}. A remotely charge compensated cubic ($O_h$) centre, an orthorhombic C$_{2v}$ symmetry centre associated with a nearest neighbour $[110]$ alkali ion, and a modified tetragonal centre having an on-axis $[200]$ alkali ion in the next nearest neighbour position.

\section{Theoretical description \label{sec:theory}}

This section details both the basic elements of the crystal field and spin Hamiltonian formalisms before describing the method employed to perform the enhanced crystal-field fit outlined in the introduction. 

\subsection{Crystal field Hamiltonian \label{subsec:cf}}

The term ``crystal-field Hamiltonian''  in the context of rare-earth doped insulators is used to describe an effective Hamiltonian for the $4f^N$ configuration, containing parameters fit to phenomenological data \cite{liu_electronic_2006,Reid2016}. The complete Hamiltonian reads 
\begin{equation}
  H = H_{\mathrm{FI}} + H_{\mathrm{CF}} + H_{\mathrm{Z}} + H_{\mathrm{HF}},
  \label{eqn:h_defn}
\end{equation}
where $H_{\mathrm{FI}}$ corresponds to the free-ion Hamiltonian, $H_{\mathrm{CF}}$ is the Hamiltonian describing the effect of the crystal on the ion, $H_{\mathrm{Z}}$ is the Zeeman Hamiltonian, and $H_{\mathrm{HF}}$ is the hyperfine Hamiltonian. 

The two dominant interactions of the free-ion Hamiltonian are the electrostatic repulsion and the spin-orbit interaction.  We follow Carnall \emph{et al.} \cite{carnall_systematic_1989} in defining the free-ion Hamiltonian by
\begin{eqnarray}
  H_{\mathrm{FI}} &=& E_{\mathrm{AVG}} + \sum_{2,4,6} F^k f_k + \zeta_{4f} A_{\mathrm{SO}} + \alpha L(L+1) + \beta G(G_2) + \gamma G(R_7)  
\nonumber \\
&+& \sum_{i = 2,3,4,6,7,8} T^i t_i
+ \sum_{h=0,2,4} M^h m_h
+ \sum_{k=2,4,6} P^k p_k , 
  \label{eqn:h_fi_defn}
\end{eqnarray}
where $E_{\mathrm{AVG}}$ is the spherically symmetric one-electron part of the Hamiltonian; $F^k$ are the Slater parameters; $\zeta_{4f}$ is the spin-orbit coupling constant; and $f_k$ and $A_{\mathrm{SO}}$ are the angular parts of the electrostatic repulsion and the spin-orbit coupling, respectively.  The constants $\alpha$, $\beta$, and $\gamma$ parametrize two-body interactions (Trees parameters) and the constants $T^i$ parametrize three-body interactions (Judd parameters).  
The  $M^h$ and  $P^k$ parameters are higher-order relativistic corrections. 

We now consider the second term in Eq.\ (\ref{eqn:h_defn}), $H_{\mathrm{CF}}$.  
For a cubic system,  $H_{\mathrm{CF}}$ is commonly written as \cite{wells_polarized_2000}: 
\begin{eqnarray}
  H_{\mathrm{cubic},4} &=&B_C^4 \bigg[C_0^{(4)} + \sqrt{\frac{5}{14}}(C_4^{(4)} + C_{-4}^{(4)})\bigg] \nonumber \\
     &+ &B_C^6 \bigg[C_0^{(6)} - \sqrt{\frac{7}{2}} (C_4^{(6)} + C_{-4}^{(6)}))\bigg].
    \label{eqn:cubic_cf_h_c4v}
\end{eqnarray} 
Here, the $B_\mathrm{C}^k$ are the crystal-field parameters for cubic-symmetry sites  and C$_{q}^{(k)}$ are spherical tensors, expressed in Wybourne's normalization \cite{wybourne_spectroscopic_1965}. The subscript ``$4$'' on $H$ indicates that the $z$ axis is a four-fold axis.

\begin{figure}[tb!]
  \centering
  \includegraphics[scale=1.5]{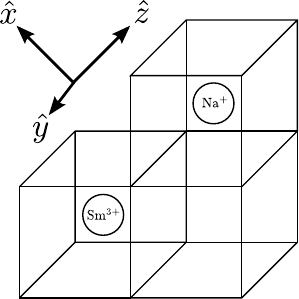}
  \caption{\label{fig:structure}%
   Axis choice used for calculations for the C$_{2v}$-oriented Sm$^{3+}$/Na$^{+}$ sites in alkaline-earth fluoride crystals. The $z$ and $x$ axes are through the edges of the F$^-$ cube surrounding the Sm$^{3+}$ ion and the $y$ axis is through the face of the cube.} 
\end{figure}

\begin{table}[tb!]
  \centering
  \caption{\label{tab:structure}
  Structure of  C$_{2v}$-oriented Ca$^{2+}$/Sr$^{2+}$ sites in alkaline-earth fluoride crystals.  $R_0$ is the distance between the cation and the F$^-$. Only two of the 12 next-nearest-neighbour cations are listed, those on the $z$ axis. }
  \begin{tabular}{lccc}
    \hline \hline
    Ion & $x/R_0$ \hspace{0.5cm}  & $y/R_0$  \hspace{0.5cm} & $z/R_0$  \\ 
    \hline

 F$^{-}$  & $0$    &  $\pm\frac{1}{\sqrt{3}}$ &$\sqrt{\frac{2}{3}}$\\
 F$^{-}$  & $\pm \sqrt{\frac{2}{3}}$    &  $\pm \frac{1}{\sqrt{3}}$ &$0$\\
 F$^{-}$  & $\mp \sqrt{\frac{2}{3}}$    &  $\pm \frac{1}{\sqrt{3}}$ &$0$\\
 F$^{-}$  & $0$    &  $\pm\frac{1}{\sqrt{3}}$ &$-\sqrt{\frac{2}{3}}$\\
 Ca$^{2+}$/Sr$^{2+}$  & $0$    &  $0$ & $\pm 2 \sqrt{\frac{2}{3}}$\\

    \hline \hline
  \end{tabular}
\end{table}

 For this form of $H_{\mathrm{CF}}$, the $x$, $y$, and $z$ axes are four-fold axes.  This is not suitable for our analysis, since a C$_{2v}$ symmetry system is best described with an axis choice where the $z$ axis is two-fold symmetric while the mirror planes are perpendicular to $x$ and $y$. The structure of such a site, with the choice that the $z$ and $x$ axes are two-fold axes (through the edge of the cube formed by the F$^-$ ligands), and the $y$ axis is a four-fold axis (through a face of the cube), is given in Fig.\ \ref{fig:structure} and Tab.\ \ref{tab:structure}. 
We may define a cubic crystal-field Hamiltonian for this geometry as follows: 
\begin{eqnarray}
  H_{\mathrm{cubic},2} =&-& \frac{1}{4}B_C^4 \bigg[C_0^{(4)} - \sqrt{10}(C_2^{(4)}+C_{-2}^{(4)}) - \frac{3}{7} \sqrt{\frac{35}{2}} (C_4^{(4)} + C_{-4}^{(4)})\bigg] 
\nonumber \\
    &-& \frac{13}{8} B_C^6 \bigg[C_0^{(6)} + \frac{\sqrt{105}}{26}(C_2^{(6)}+C_{-2}^{(6)}) 
\nonumber \\
& &\qquad\qquad - \frac{5}{13} \sqrt{\frac{7}{2}}
  (C_4^{(6)} + C_{-4}^{(6)}) + \frac{\sqrt{231}}{26}(C_6^{(6)} + C_{-6}^{(6)})\bigg].
    \label{eqn:cubic_cf_h}
\end{eqnarray} 
 The subscript ``$2$'' on $H$ indicates that the $z$ axis is a two-fold axis. 
 The signs and normalizations were chosen so that the crystal-field parameters from Ref.\ \cite{wells_polarized_2000} could be used without modification. 

The local structure of the  C$_{2v}$ sites is modified by the replacement of a Ca$^{2+}$ or Sr$^{2+}$ ion by an Na$^+$ or Li$^+$, and thus the Sm$^{3+}$-Na$^+$/Li$^+$ direction defines the C$_{2v}$ symmetry axis. For the geometry of Tab.\ \ref{tab:structure}, this means that the substitution is for a Ca$^{2+}$ or Sr$^{2+}$ on the $z$ axis.  
In Sec.\ \ref{subsec:epr} we will discuss the EPR data using the knowledge that the above Hamiltonian requires that the two-fold axis, $z$, is along the $[110]$ direction (prior to any distortion caused by the substitution), and that the $x$ and $y$ axes are along the $[1\bar{1}0]$ and $[001]$ directions, respectively (so that $y$ is a four-fold axis).

We expect the C$_{2v}$ symmetry sites to only have small distortions from the cubic sites due the substitution of a Na$^+$ or Li$^+$ ion.  We therefore write the  C$_{2v}$ crystal field Hamiltonian in terms of the above cubic crystal field Hamiltonian and \emph{changes} from the cubic parameters: 
\begin{eqnarray}
  H_{\mathrm{CF}} &=& H_{\mathrm{cubic},2} \nonumber \\
  &+&  \Delta B_0^2 C_0^2 +  \Delta B_2^2(C_2^{(2)} + C_{-2}^{(2)}) \nonumber \\ 
  &+& \Delta B_0^4 C_0^{(4)} +  \Delta B_2^4(C_2^{(4)} + C_{-2}^{(4)}) +  \Delta B_4^4(C_4^{(4)} + C_{-4}^{(4)}) \nonumber \\
  &+&  \Delta B^6_0 C^{(6)}_0 +  \Delta B^6_2 (C^{(6)}_2 + C^{(6)}_{-2}) 
\nonumber \\
&+& \Delta  B^6_4 (C^{(6)}_4 + C^{(6)}_{-4}) +  \Delta B^6_6 (C^{(6)}_6 + C^{(6)}_{-6}). 
  \label{eqn:c2v_cf_h}
\end{eqnarray}
Here  $\Delta B_q^k$ is the \emph{change} in the crystal-field parameter from the cubic value. We fix $B^4_C$ and $B^6_C$ to the values obtained by fitting to the purely cubic centers \cite{wells_polarized_2000}, and then vary all of the  $\Delta B_q^k$ parameters of Eq.\ (\ref{eqn:c2v_cf_h}), which represent the distortion from cubic introduced by codoping with Na$^+$ or Li$^+$ ions. 

The Zeeman term $H_{\mathrm{Z}}$ describes the effect of  an external magnetic field. We  define \cite{judd_operator_1963}
\begin{equation}
  H_{\mathrm{Z}} = \mu_{\mathrm{B}} \sum_i^N \mathbf{B} \cdot (\mathbf{l}_i + 2 \mathbf{s}_i),
  \label{eqn:h_Z_defn}
\end{equation}
where $\mu_{\mathrm{B}}$ is the Bohr magneton, and $\mathbf{l}_i$ and $\mathbf{s}_i$ are the orbital and spin angular momenta of the $i$th electron for $i$ in the $f^N$ configuration, respectively. 

 The final term in Eq.\ (\ref{eqn:h_defn}), $H_{\mathrm{HF}}$, describes the coupling of nuclear spin with the spin of the $4f^N$ electrons. 
Following Judd \cite{judd_operator_1963}, the hyperfine Hamiltonian is 
\begin{equation}
  H_{\mathrm{HF}} = a_l \sum_{i=1}^N \mathbf{N}_i \cdot \mathbf{I},
  \label{eqn:h_hf_defn}
\end{equation}
with
\begin{equation}
  \mathbf{N}_i = \mathbf{l}_i - \mathbf{s}_i + \frac{3 \mathbf{r}_i (\mathbf{s}_i \cdot \mathbf{r}_i)}{r_i^2},
  \label{eqn:N_i_defn}
\end{equation}
where $\mathbf{r}_i$ is the position of the $i$th $4f$ electron.  Furthermore, $\mathbf{I}$ is the total angular momentum of the nucleus, and the constant factor $a_l$ is defined by
\begin{equation}
  a_l = \frac{2 \mu_{\mathrm{B}} \mu_{\mathrm{N}} g_N}{\langle r_i^3 \rangle},
  \label{eqn:a_l_defn}
\end{equation}
where $\mu_{\mathrm{N}}$ is the nuclear magneton, $g_N$ is the nuclear $g$-factor and $\langle r_i^3 \rangle$ is the mean cube radius of the $4f$ orbital. 
For details on how the hyperfine interaction is treated in terms of tensor operators, the reader is referred to Refs.\ \cite{mcleod_intensities_1997,wells2004hyperfine}

Having defined the complete Hamiltonian, the energies and wavefunctions can be calculated by diagonalizing $H$ in the $\ket{\gamma L S J M_J I M_I}$ basis.  Here the state label $\gamma$ corresponds to all additional degrees of freedom of the electronic configuration.  The reader is referred to the monographs of Wybourne \cite{wybourne_spectroscopic_1965} and Judd \cite{judd_operator_1963} for a detailed discussion of these quantum numbers.

\subsection{Superposition Model \label{subsec:superposition}}

The superposition model, developed by Newman and co-workers in the late 60s
\cite{New71,NN89a,NeNg00}, has proved an important framework for the
understanding of crystal-field  parameters. The model uses the approximation
that the contributions to the effective potential from different ligands is
superposeable (for an electrostatic potential this is obviously exact). If we
had a single ligand $L$ on the $Z$ axis at distance $R_0$ the resulting
cylindrical symmetry would restrict only crystal-field parameters with $q=0$ to
be non-zero. We define the \emph{intrinsic} parameters for this ligand as
$\bar{B}_k(L, R_0)$ to be $B^{(k)}_0$ parameters for this single ligand. Adding
together the effects of all the ligands yields
\begin{equation} \label{eq:sm}
B^{k}_q = 
 \sum_L \bar{B}_k(L,R_0) (-1)^q C^{(k)}_{-q}(\theta_L, \phi_L)
 \left(\frac{R_0}{R_L}\right)^{t_k}. 
\end{equation}
In this equation, $(-1)^q C^{(k)}_{-q}(\theta_L, \phi_L)$ is required to take into
account the effect of rotating a ligand from the $Z$ axis to orientation
$(\theta_L, \phi_L)$. This angular term has the same form as in a
point-charge model. The term $\left({R_0}/{R_L}\right)^{t_k}$, which
takes into account the variation of the interactions with distance,
would be the same as for a point-charge potential if we chose $t_k =
k+1$. However, analyses of experimental crystal-field parameters suggest
that the power law is generally higher than electrostatic power
laws, because the quantum-mechanical effects that give rise to
the majority of the ``crystal field'' involves overlap of ligand
orbitals, which falls off faster than electrostatic potentials. 

Equation (\ref{eq:sm}) may be used to demonstrate that the parametrization of Eqn.\ (\ref{eqn:c2v_cf_h}) is consistent with geometry of the nearest-neighbor F$^-$ ions given in Tab.\ \ref{tab:structure}.

\subsection{Spin Hamiltonian \label{subsec:sh}}
Most studies of magnetic and hyperfine interactions make use of a spin Hamiltonian.  It is defined as \cite{macfarlane_coherent_1987} 
\begin{equation}
  \mathscr{H} = [\mathscr{H}_{\mathrm{FI}} + \mathscr{H}_{\mathrm{CF}}] + [\mathscr{H}_{\mathrm{HF}} + \mathscr{H}_{\mathrm{Z}}].
  \label{eqn:sh}
\end{equation}
The first two bracketed terms account for free-ion and crystal-field interactions as defined by Eqs.\ (\ref{eqn:h_fi_defn}) and (\ref{eqn:c2v_cf_h}), respectively.  Provided the state under consideration is sufficiently separated from adjacent states, the terms $\mathscr{H}_{\mathrm{FI}}$ and $\mathscr{H}_{\mathrm{CF}}$ amount to a constant off-set due to the spin-orbit and crystal-field interactions.  The hyperfine structure of the state in the presence of a magnetic field can then be accurately characterized by the interactions in the second set of parentheses.  These terms are the nuclear hyperfine interaction and the electronic Zeeman interaction \cite{macfarlane_coherent_1987}.  We note that contributions due to the nuclear quadrupole as well as nuclear Zeeman interactions have been neglected here.  The effective Hamiltonian $\mathscr{H}$ for a specific electronic state is referred to as the spin Hamiltonian. The details of the components that contribute to the matrix elements of these interactions depends on whether one is dealing with a Kramers or non-Kramers ion, as well as the local site symmetry.

In the case of samarium in a C$_{2v}$ symmetry center, the appropriate spin Hamiltonian has the form
\begin{equation}
  \mathscr{H} = \mu_B \mathbf{B} \cdot \mathbf{g} \cdot \mathbf{S} + \mathbf{I} \cdot \mathbf{A} \cdot \mathbf{S}.
  \label{eqn:kramers_sh}
\end{equation}
If $\mathbf{g}$ is symmetric (which is the normal choice), the $g$-tensor can be diagonalized by an Euler rotation $R$ according to
\begin{equation}
  \mathbf{g} = R(\alpha_g, \beta_g, \gamma_g)
  \left( 
  \begin{array}{ccc}
    g_x & 0 & 0 \\
    0 & g_y & 0 \\
    0 & 0 & g_z
    \end{array} 
  \right)
            R^{\mathrm{T}}(\alpha_g, \beta_g, \gamma_g),
  \label{eqn:g_ten_defn}
\end{equation}
with principal $g$-values $g_x$, $g_y$, and $g_z$ and Euler rotation parameters $\alpha_g$, $\beta_g$, and $\gamma_g$.  The orientation for which the $g$-tensor is diagonal is referred to as the principal axes of the Zeeman term.  Similarly, the hyperfine parameter matrix $\mathbf{A}$ can be diagonalized by an Euler rotation to their respective principal axes. 
We shall see below that our choice of C$_{2v}$ axes automatically diagonalises the $g$ tensor, considerably simplifying the analysis.

\subsection{Crystal-field analysis \label{subsec:cf_analysis}}

In order to fit the parameters of Eqs.\ (\ref{eqn:h_fi_defn}) and (\ref{eqn:c2v_cf_h}) to both energy level and spin Hamiltonian data we must calculate the theoretical spin Hamiltonian for a given set of free-ion and crystal-field parameters. Consequently, we must calculate the theoretical spin-Hamiltonian parameters for a given $4f^N$ configuration Hamiltonian, $H$. This projection can be cast in the language of effective operators \cite{hurtubise_algebra_1993,hurtubise_algebra_1993-1}.  We proceed by defining the $4f^N$ configuration as the complete space and the spin Hamiltonian basis as the model space.  Then, if we have eigenstates of the complete Hamiltonian of the form 
\begin{equation*}
  H \ket{\lambda} = E_\lambda \ket{\lambda},
\end{equation*}
we require the matrix elements of the model space Hamiltonian, such that 
\begin{equation*}
  \mathscr{H} \ket{\lambda_0} = E_\lambda \ket{\lambda_0},
\end{equation*}
where $\mathscr{H}$ is the spin Hamiltonian defined by Eq.\ (\ref{eqn:sh}).  Since the spin Hamiltonian method assumes that $H_{\mathrm{FI}}$ and $H_{\mathrm{CF}}$ can be considered as a zero-order Hamiltonian, the transformation to the spin Hamiltonian basis can be performed by diagonalizing $H_{\mathrm{FI}} + H_{\mathrm{CF}}$ and then, given the resulting eigenvectors $V$, one can transform the Zeeman and hyperfine operators according to
\begin{equation}
  A_{\mathrm{eff}} = V^\dag A V,
  \label{eqn:operator_proj}
\end{equation}
for operator, $A$, and effective operator, $A_{\mathrm{eff}}$.

\section{Experimental techniques \label{sec:exp_tech}}

The samples of CaF$_2$ or SrF$_2$ codoped with either SmF$_3$ with LiF or SmF$_3$ with NaF were prepared using the vertical Bridgman-Stockbarger method.  The growth was performed in a positive argon atmosphere using a 38 kW Arthur D. Little radio frequency furnace.  The resulting crystal boules were oriented along the $[100]$ axis using Laue x-ray diffraction, and cut into 5 mm$^3$ samples which were further polished for optical spectroscopy.  

A series of crystals with various dopant concentrations were prepared from which the following were chosen for detailed study: two crystals of 0.075\%Sm$^{3+}$:1.09\%Na$^+$:CaF$_2$ and 0.026\%Sm$^{3+}$:0.99\%Li$^{+}$:CaF$_2$, in addition to two crystals of 0.01\%Sm$^{3+}$:0.86\%Na$^+$:SrF$_2$ and 0.037\%Sm$^{3+}$:1.17\%Li$^{+}$:SrF$_2$. These dopant concentrations were selected since they yielded primarily single-ion centers for crystals prepared without the NaF or LiF codopant \cite{wells_polarized_2000}. We note that in the case of CaF$_2$ preferentially formed clusters will be present at the selected dopant levels. 

The Sm$^{3+}$ ions were excited for both excitation and fluorescence spectroscopy using a Spectra Physics 375B dye laser, employing Rhodamine 560 dye dissolved in ethylene-glycol, pumped by a 5 W Coherent Innova 70 argon-ion laser. The samples were cooled to 10 K using a closed-cycle CTI-Cryogenics model 22C cryostat. A Spex Industries 500M single monochromator was employed for dispersing fluorescence. Dispersed light in the 18,000 cm$^{-1}$ to 12,000 cm$^{-1}$ wavelength range was then detected using a thermoelectrically cooled Hamamatsu R9249 photomultipler. For fluorescence in the infrared, phase sensitive detection was performed with a liquid nitrogen cooled germanium detector in conjunction with an Ortholoc model 9502 lock-in amplifier. 

The electron-paramagnetic resonance (EPR) measurements were performed at temperatures between 5-50 K using a Bruker EMX10/12 X-band spectrometer with microwave frequencies in the range 9.695-9.710 GHz, a microwave power of 0.1 mW, and 100 kHz field modulation.  The angular variations of the EPR spectra were measured by rotating the sample in the cavity. The full range of the applied magnetic field was between 0-1.5 T.

\section{Results and discussion \label{sec:results}}

\subsection{Energy levels of Sm$^{3+}$ ions}

The valence electrons of Sm$^{3+}$ are of the $4f^5$ configuration, which  consists of a total of 1001 Kramers doublets.  Fluorescence in the visible region was detected from the $^{4}${G}$_{5/2}$ multiplet to all $^{6}${H}$_{J}$ and $^{6}${F}$_{J}$ multiplets, with the exception of the $^{6}${F}$_{11/2}$.  

The standard notation of Dieke to label specific crystal-field levels is used throughout this communication. Specifically, the ground multiplet is labelled as $Z$, with the ground electronic state designated as $Z_1$. Further, the first excited multiplet is labelled by the letter $Y$, and the $^{4}${G}$_{5/2}$ multiplet is labelled by the letter $A$. Again, numerical subscripts are used to indicate a specific electronic state. For the $^{4}${F}$_{1/2}$, $^{6}${F}$_{3/2}$, and $^{6}${H}$_{13/2}$ multiplets we adopt the notation employed in our previous work in which we opted to label them by the letter $S$ since crystal-field $J$-mixing means the corresponding wavefunctions of these multiplets are heavily mixed.

\subsection{Laser spectroscopy \label{subsec:laser_spect}}

\begin{figure}[tb!]
  \centering
  \includegraphics[scale=0.55]{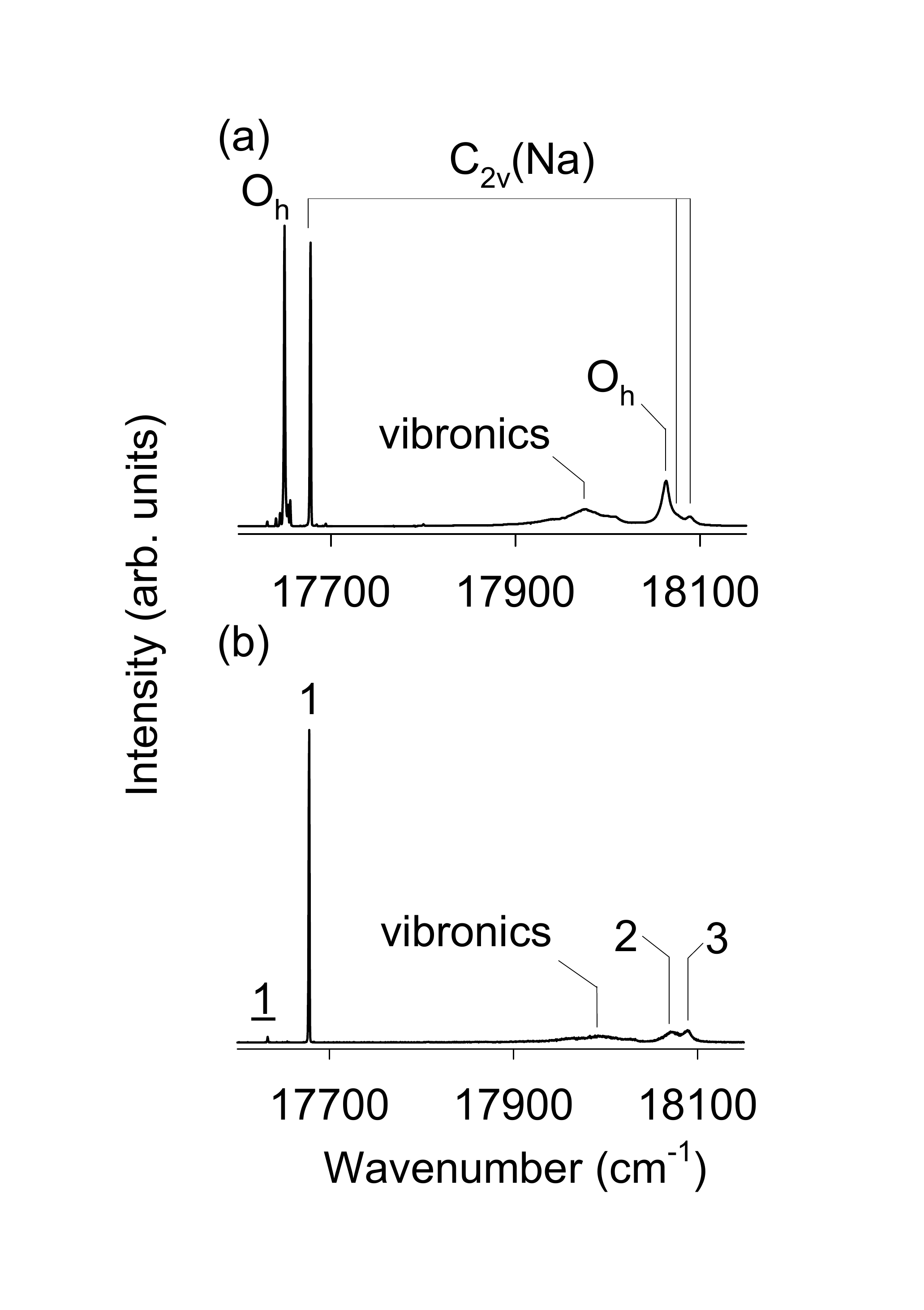}
  \caption{\label{fig:smnacaf2_excitation}%
  10 K excitation spectra of the $^{4}${G}$_{5/2}$ multiplet of Sm$^{3+}$:Na$^+$:CaF$_2$. (a) broadband excitation spectrum of both the cubic and C$_{2v}$ centers; (b) site-selective excitation spectrum of the  C$_{2v}$ center -- here $\underline{1}$ denotes the Z$_2 \to$ A$_1$ transition.} 
\end{figure}

Site-selective and broadband excitation spectroscopy was performed on all four materials. 
A representative broadband excitation spectrum for 0.075\%Sm$^{3+}$:1.09\%Na$^+$:CaF$_2$ recorded at a temperature of 10 K is shown in Figure \ref{fig:smnacaf2_excitation} (a). Both cubic and C$_{2v}$ centres have been identified; the cubic centre has been investigated in detail in the past \cite{wells_polarized_2000}.  We will henceforth denote the orthorhombic symmetry centre by C$_{2v}$(Na$^+$/Li$^{+}$). 
Additionally, we observe a phonon sideband, shifted by an energy of $\sim 350$ cm$^{-1}$ from the zero-phonon line. Such sidebands are a common feature of this multiplet. Figure \ref{fig:smnacaf2_excitation} (b) consists of an excitation spectrum selecting only the C$_{2v}$(Na$^+$) center. Fluorescence was recorded by monitoring the $A_1 \to Y_1$ transition at an energy of 17622 cm$^{-1}$. We note that the 17 cm$^{-1}$ splitting of the $^4\Gamma_8$ level (lines 2 and 3 in Figure \ref{fig:smnacaf2_excitation} (b))  suggests that this center is very close to cubic symmetry. This is also reflected in the non-cubic crystal-field parameters presented in Tab.\ \ref{tab:smca2_srf2_params}. 

The phonon sideband is due to an electron interaction with local vibronic modes around a substitutional Sm$^{3+}$ ion.  Similar phonon sidebands have been observed for Ce$^{3+}$:Na$^+$:CaF$_2$(SrF$_2$) and were located in the range from 100 to 500 cm$^{-1}$ shifted from the zero-phonon lines of Ce$^{3+}$ \cite{manthey_crystal_1973,pack_ce3+:na+_1989}.  In general, the difference in the relative intensities of the phonon sideband to the zero-phonon line is due to the electron-phonon coupling strength. The larger coupling constant corresponds to a larger overlap of the wavefunctions between the central Sm$^{3+}$ (Ce$^{3+}$) ion and the eight ligand F$^-$ ions in CaF$_2$.  This value is estimated from the superhyperfine (SHF) interaction of the central ion and the ligand ions in the EPR measurement (Fig.\ \ref{fig:smnacaf2_shf}).

\begin{figure}[tb!]
  \centering
  \includegraphics[scale=0.30]{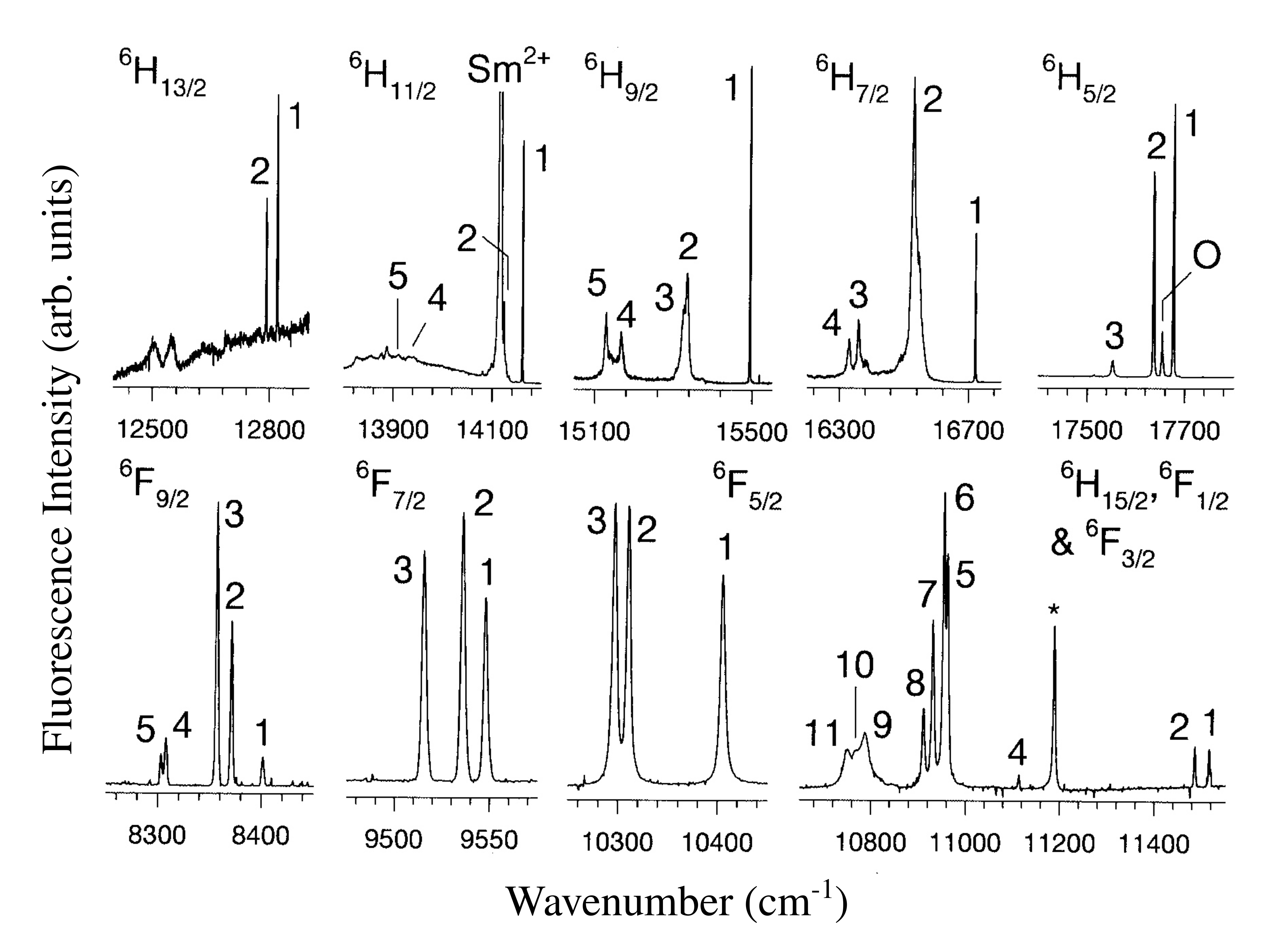}
  \caption{\label{fig:site_selective_excitation}%
  10 K fluorescence of the C$_{2v}$(Na$^+$) centre in Sm$^{3+}$:Na$^+$:CaF$_2$. Numbered peaks are assigned to the C$_{2v}$ centre. Other peaks are explained in the text.} 
\end{figure}

Figure \ref{fig:site_selective_excitation} shows representative site-selective fluorescence spectra for the C$_{2v}$(Na$^+$) center in 0.075\%Sm$^{3+}$:1.09\%Na$^+$:CaF$_2$ as measured at 10 K. The spectrum measured for the $^{4}${G}$_{5/2}$ $\to$ $^{6}${H}$_{5/2}$ transitions was recorded using excitation of the Z$_1 \to \mathrm{A}_3$ transition. This necessarily excites the cubic center as well, with the cubic A$_1 \to \mathrm{Z}_1$ transition observed at 17653 cm$^{-1}$. The fluorescence spectra of all other multiplets were obtained exciting the C$_{2v}$(Na$^+$) Z$_1 \to \mathrm{A}_1$ transition at 17678 cm$^{-1}$. The spectra of the $^{6}${H}$_{9/2}$ and $^{6}${H}$_{13/2}$ multiplets is partially obscured by the unintentional presence of Sm$^{2+}$ produced in the reducing conditions during crystal growth. A similar effect is observed for Li$^{+}$ co-doped CaF$_2$ crystals although no Sm$^{2+}$ was present in the SrF$_2$ samples studied. A strong Sm$^{2+}$ peak near 14118 cm$^{-1}$ with a broad sideband is observed corresponding to the single electric dipole allowed $5d \to 4f$  transition along with a vibrational sideband \cite{wood_absorption_1962}. The peak observed at 11189 cm$^{-1}$ is present in both Sm$^{3+}$:Li$^{+}$:CaF$_2$ and Sm$^{3+}$:Na$^+$:CaF$_2$ and is, as such, unassigned. Detailed energy level assignments for $^{6}${H} and $^{6}${F} terms as well as the $^{4}${G}$_{5/2}$ multiplet of the C$_{2v}$ centers in Sm$^{3+}$:Na$^+$/Li$^{+}$:CaF$_2$ and Sm$^{3+}$:Na$^+$/Li$^{+}$:SrF$_2$ are presented in Tab.\ \ref{tab:eval}. 

\subsection{Electron-paramagnetic resonance \label{subsec:epr}}

\begin{figure}[tb!]
  \centering
  \includegraphics[scale=0.55]{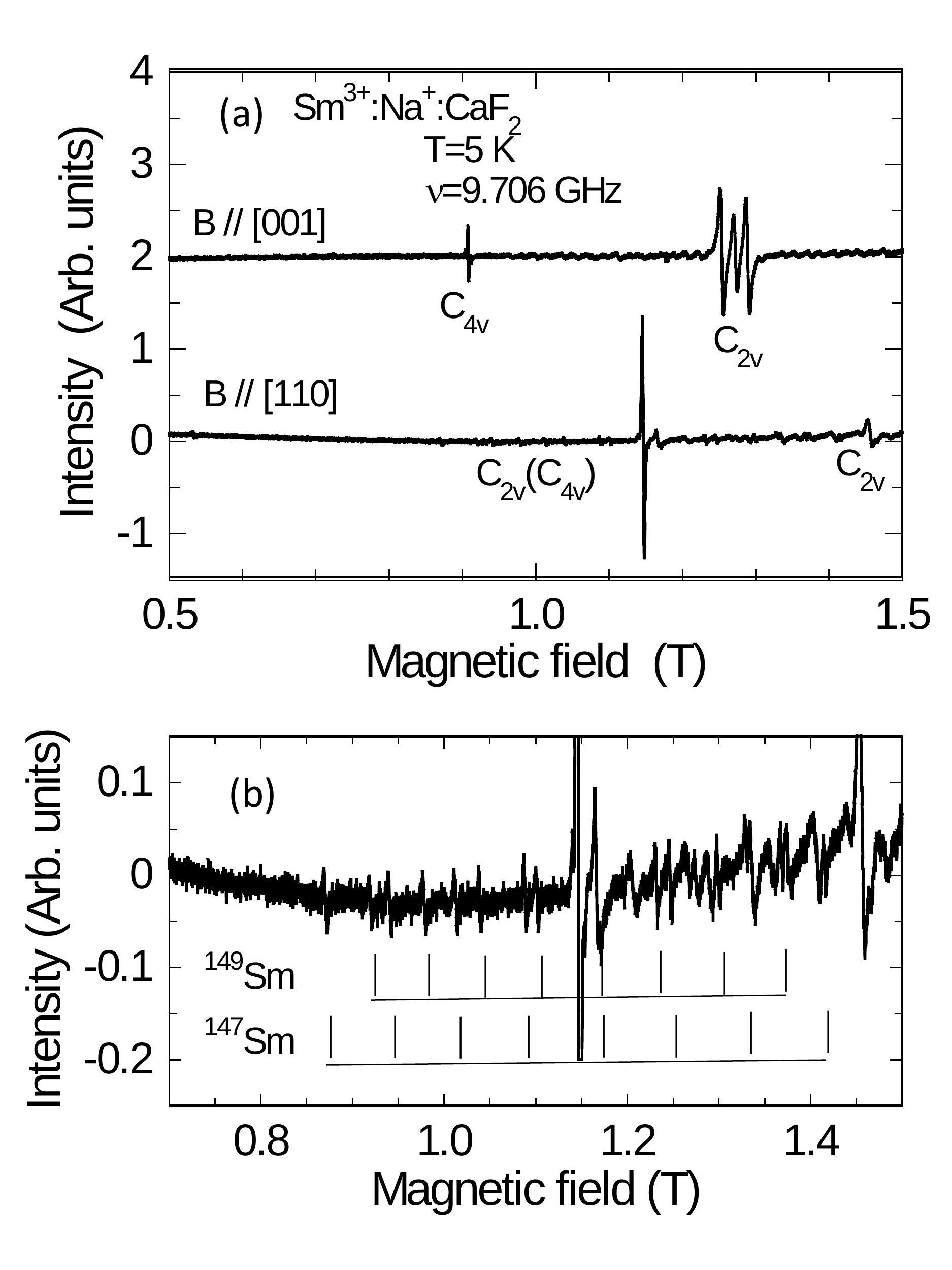}
  \caption{\label{fig:smnacaf2_esr}%
  (a) The EPR spectra for Sm$^{3+}$:Na$^+$:CaF$_2$ with $B$ parallel to the $[001]$ and $[110]$ directions, at 5 K.  The labels of C$_{4v}$ and C$_{2v}$ denote tetragonal and orthorhombic centres, respectively.
(b) The expanded $B$ parallel to $[110]$ spectrum consists of two sets of hyperfine structure lines due to $^{147}$Sm and  $^{149}$Sm isotopes.} 
\end{figure}

Figure \ref{fig:smnacaf2_esr}(a) shows a typical Sm$^{3+}$ EPR spectrum measured for Sm$^{3+}$:Na$^+$:CaF$_2$ with $\mathbf{B} \parallel$ $[001]$ and $[110]$ at 5 K. The spectra consist of two groups of intense EPR lines, denoted by $C_{4v}$ and $C_{2v}$, due to the $^{152}$Sm isotopes with $I=0$ and natural abundance of 26.8\%.  Two sets of weak octet hyperfine lines in Figure \ref{fig:smnacaf2_esr}(b) are due to the $^{147}$Sm and $^{149}$Sm isotopes with a non-zero nuclear spin of $I=7/2$ and natural abundances of 15.1\% and 13.8\%, respectively. The hyperfine coupling constants are estimated from the magnetic field separation between adjacent lines using second-order perturbation theory. 
The coupling constants reported in Tab.\ \ref{tab:smcaf2srf2_shparam} are the average for $^{147}$Sm and $^{149}$Sm. 

\begin{figure}[tb!]
  \centering
  \includegraphics[scale=0.6]{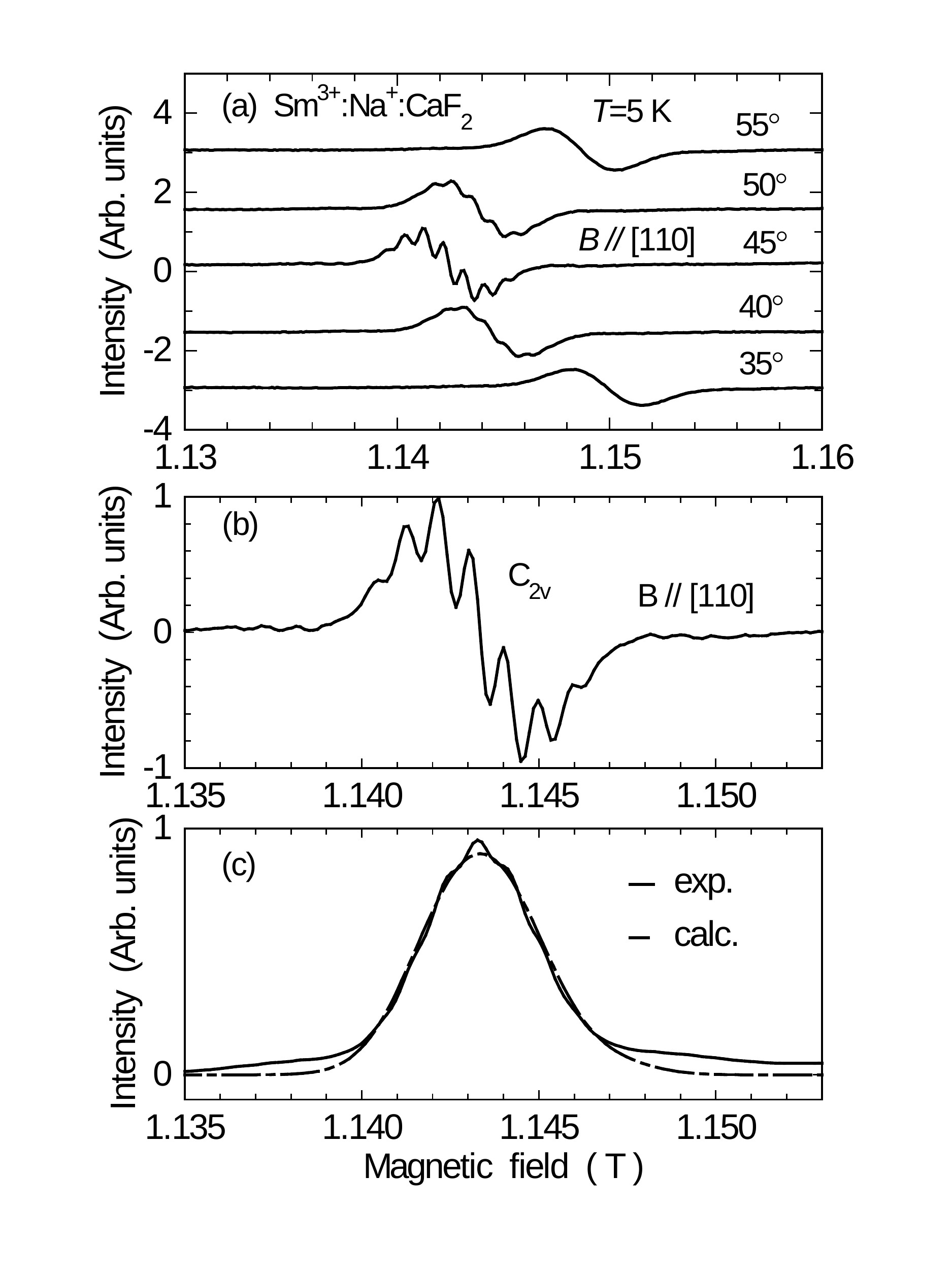}
  \caption{\label{fig:smnacaf2_shf}%
  (a) Angular variations of the EPR spectra around the $[110]$ direction in the $\{001\}$ plane observed in Sm$^{3+}$:Na$^+$:CaF$_2$ at 5\,K. 
  (b) The superhyperfine structure of the orthorhombic centre observed when the magnetic field was applied to $[110]$ in Sm$^{3+}$:Na$^+$:CaF$_2$. (c) The integrated EPR line is in agreement with the sum of nine Gaussian lines with the intensity ratio of 1:8:28:56:70:56:28:8:1, assuming an isotropic SHF interaction of the central samarium ion with the eight equivalent fluorine ligand ions having a total nuclear spin of 4 $(=8 \times 1/2)$. 
 } 
\end{figure}

Figure \ref{fig:smnacaf2_shf}(a) shows the angular variation of the EPR spectrum around the $[110]$ direction in the $\{001\}$ plane at 5\,K. There are six geometrically equivalent  C$_{2v}$ Sm$^{3+}$ centres, with principal axes along the $[110]$, $[1\bar{1}0]$, $[011]$, $[01\bar{1}]$, $[101]$, and $[10\bar{1}]$ directions. The line at 1.143\,T with B parallel to $[110]$ is a unique line from the single nondegenerate centre with the principal axis in the $[110]$ direction. 

The EPR line shape for any magnetic field direction (see the spectra at $35^\circ$ and $55^\circ$ in Fig.\ \ref{fig:smnacaf2_shf}(a)) is Gaussian and the linewidth is created by the superhyperfine (SHF) interaction of Sm$^{3+}$ with the surrounding F$^-$ ions. 
However, as shown in  Figure \ref{fig:smnacaf2_shf}(a), the SHF structure only appears when the magnetic field is parallel to the $[110]$ direction. The number of SHF lines is clearly more than seven, as shown in  Figure \ref{fig:smnacaf2_shf}(b), suggesting that the total spin contributing to this structure must be greater than 3. Since the nuclear spin of an F$^-$ ion is $\frac{1}{2}$ this requires more than 6 F$^-$ ions. We therefore consider the eight equivalent nearest-neighbour F$^-$ ligands having a total nuclear spin of $4$ ($=8\times\frac{1}{2}$). The nine SHF lines are calculated to have the intensity ratio of 1:8:28:56:70:56:28:8:1. 
The calculated curve fitted to the observed integrated EPR line is shown in Fig.\ \ref{fig:smnacaf2_shf}(c). This result suggests an isotropic SHF interaction of the central samarium ion with the eight equivalent fluorine ligand ions.
The resolved SHF lines of Sm$^{3+}$ with eight equivalent F$^-$ ligands are not observed around 1.28\,T with the magnetic field parallel to the $[001]$ direction in Figure \ref{fig:smnacaf2_esr}(a). A possible explanation is that the SHF structure from the four sites with principal axes along $[011]$, $[01\bar{1}]$, $[101]$, and $[10\bar{1}]$ cancel when $B$ is parallel to $[001]$. 
The same nine resolved SHF lines are observed for SrF$_2$:Sm$^{3+}$:Na$^+$ with $B$ parallel to $[111]$, but not with $B$ parallel to $[110]$. These results are explained as follows: (i) isotropic SHF interaction is dominant, (ii) although the contribution of the $p$ orbital of the F$^-$ ligands to the SHF interaction is small compared with that of the $s$ orbital, it produces an anisotropic SHF splitting for any magnetic field. This is reflected by the unexpected appearance of the resolved SHF structure. Here we consider the dominant SHF interaction represented by the Fermi contact term.   
The coupling constant, $A_{\mathrm{SHF}}$, is given by
\begin{equation}
  A_{\mathrm{SHF}} = \frac{8}{3} \pi g_e \mu_B g'_N \mu_N | \Psi(0) |^2 f_s,
  \label{eqn:ASHF}
\end{equation}
where $g_e$ is the $g$-value of the Sm$^{3+}$ ground-state electron spin, $g'_N$ is the $g$-value of the F$^-$ ligand nuclear spin, $|\Psi(0)|^2$ is the electron density at the F$^-$ nucleus \cite{schoemaker_111-oriented_1966}, and $f_s$ is the s-character amount of the F$^-$ ligand ion electron.
The field separation, $\Delta B$, between the adjacent lines of the nine SHF lines is experimentally estimated to be 0.9 mT as shown in Fig.\ \ref{fig:smnacaf2_shf}(b). The relation between the SHF coupling constant and the separation magnetic field is given by
\begin{equation}
  A_{\mathrm{SHF}} = g_e \mu_B \Delta B.
  \label{eqn:Ashf}
\end{equation}
 Using Eqs.\ (\ref{eqn:ASHF}) and (\ref{eqn:Ashf}), with a $g$-value for Sm$^{3+}$ of $g_e=0.606$, $g'_N=5.254$ of F$^-$ and $|\Psi(0)|^2=78\times 10^{24}$ cm$^{-3}$ for the F$^-$ ligand ion \cite{schoemaker_111-oriented_1966}, the value of $f_s$ is calculated to be 0.05\%.  This value is very small, but not negligible.  Such an expansion of the electron wavefunction of the Sm$^{3+}$ central ion towards the F$^-$ ligand ions is indicative of strong electron-phonon coupling, which results in the observed phonon-side bands shown in Fig.\ \ref{fig:smnacaf2_excitation}.  

\begin{figure}[tb!]
  \centering
  \includegraphics[scale=0.65]{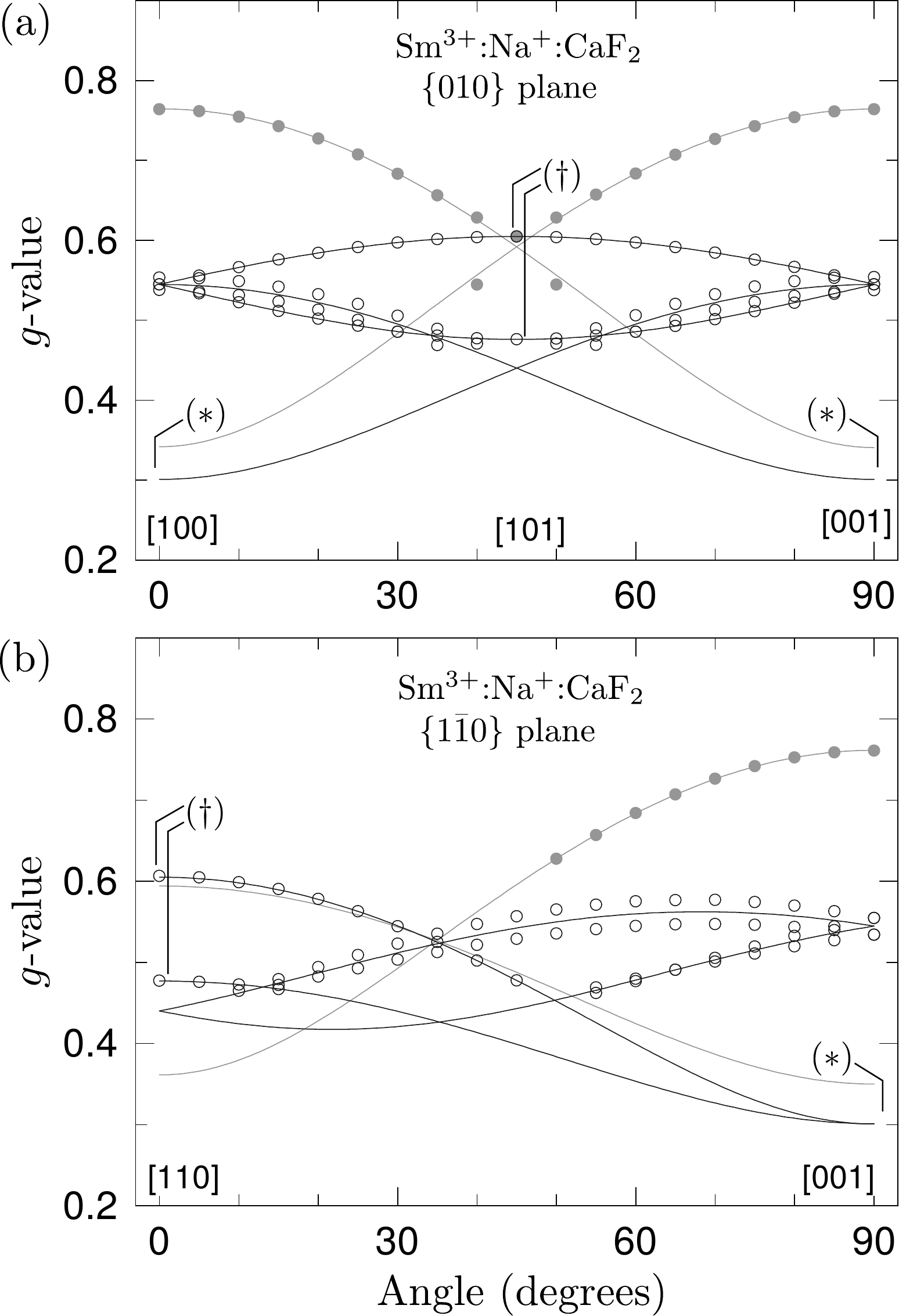}
  \caption{\label{fig:smnacaf2_esr_ang}%
  The angular variations of the EPR lines ($I=0$) observed in Sm$^{3+}$:Na$^+$:CaF$_2$ with T=5 K in (a) the $\{010\}$ plane and (b) the $\{1 \bar{1} 0\}$ plane.  The resonance lines assigned to the C$_{4v}$ and C$_{2v}$ sites are denoted by solid and open circles, respectively. For the C$_{2v}$ site the two extrema assigned to  $g_x$ and $g_z$ are indicated by ($\dag$), and the extremum assigned to  $g_y$ by ($*$).
}
\end{figure}

Figure \ref{fig:smnacaf2_esr_ang} shows the angular variations of the $g$-values in the $\{010\}$ and $\{110\}$ planes for the $^{152}$Sm$^{3+}$ ion with zero nuclear spin in Sm$^{3+}$:Na$^+$:CaF$_2$.  As the full magnetic field range is 0-1.5 T, the $g$-values below 0.46  were not observable.  The curves in Fig.\ \ref{fig:smnacaf2_esr_ang} are calculated using the spin Hamiltonian (\ref{eqn:kramers_sh}) without the hyperfine interaction. The patterns of the angular variations show tetragonal and orthorhombic symmetries. In Ref.\ \cite{yamaga_role_2002}, the principal $x$, $y$ and $z$ axes of the C$_{4v}$ centres are defined as the $[100]$, $[010]$ and $[001]$ directions, respectively. The C$_{2v}$ symmetry axes must be consistent with the definition of our crystal-field Hamiltonian (\ref{eqn:c2v_cf_h}). As noted earlier, the reduction from cubic symmetry is achieved by the replacement of a Ca$^{2+}$ ion by an Na$^+$ or Li$^+$ ion, and the C$_{2v}$ symmetry axis is therefore along the Sm$^{3+}$-Na$^+$/Li$^+$ direction. Observing that the C$_{2v}$ axis necessarily intersects an edge of the cube formed by the 8 nearest-neighbor F$^-$ ions, it follows that one of the remaining two axes also passes through a fluorine cube edge, while the other axis must intersect a fluorine cube face. From Fig.\ \ref{fig:smnacaf2_esr_ang}(a) it is apparent that there are two extrema in the $[110]$ direction (indicated by ($\dag$)), and one extremum in the $[001]$ direction (equivalent to the $[101]$ direction in the figure, indicated by ($*$)). The $g_z$ extremum must be along $[110]$ direction.  One of  $g_x$ or $g_y$ must be along $[110]$ and the other along $[001]$, which has a unique extremum. We make the following axis identifications: the principal $z$ axis corresponds to the $[110]$ direction, while the $x$ and $y$ directions are in the $[1\bar{1}0]$ and $[001]$ directions, respectively, consistent with Tab.\ \ref{tab:structure} and Eq.\  \ref{eqn:cubic_cf_h}. Note that the cubic crystal structure means that there will be sites with the local $x$, $y$ and $z$ axes along different crystal directions, which explains why there are two extrema along $[110]$. The distinction between these two extrema is given by our calculation, with $g_x = 0.606$ being greater than $g_z = 0.477$, whereas the unique extrema is $g_y=0.3$.  

The cubic (O$_h$) symmetry centers, observed in the laser selective excitation measurements in Fig.\ \ref{fig:smnacaf2_excitation}, could not be detected using EPR. We tentatively ascribe this to fast spin-lattice relaxation between the magnetically split ground state ($Z_1$ of $^{6}${H}$_{5/2}$) and the consequently broadened EPR linewidths. Analogous measurements were made for Sm$^{3+}$:Li$^+$:CaF$_2$, Sm$^{3+}$:Li$^+$:SrF$_2$, and Sm$^{3+}$:Na$^+$:SrF$_2$, as discussed below.

Table \ref{tab:smcaf2srf2_shparam} summarizes the $g$-values and hyperfine coupling constants, $A$, obtained by fitting the spin Hamiltonian (\ref{eqn:kramers_sh}) to the recorded data.  This is displayed alongside the spin Hamiltonian parameters inferred from the crystal-field Hamiltonian, as will be discussed in Sec.\ \ref{sec:cf_fit}.

The orthorhombic C$_{2v}$ symmetry centers are strongly associated with Li$^{+}$ and Na$^{+}$ charge compensators.  The principal $z$ axis of the Li$^+$ orthorhombic center is tilted by an angle of about $12^\circ$ from the $[110]$ direction, whereas that of the orthorhombic C$_{2v}$(Na$^+$) center is parallel to the $[110]$ direction.  This distinction is due to the differing ionic radii of Li$^+$ (0.088 nm) and Na$^+$ (0.130 nm).  As the ionic radius of Na$^+$ is closer to those of Ca$^{2+}$ (0.126 nm) and Sr$^{2+}$ (0.139 nm), Na$^+$ ions substitute for Ca$^{2+}$ with minimal relaxation of the surrounding ions. On the other hand, Li$^+$ ions are smaller in size, presumably leading to two minima of the chemical potential with the Li$^+$ ions offset from the substitutional position.

\subsection{Crystal-field fit \label{sec:cf_fit}}

In this section, we outline the procedure used to fit the crystal-field parameters for the C$_{2v}$ centres in CaF$_2$:Sm$^{3+}$:Na$^+$/Li$^{+}$ and SrF$_2$:Sm$^{3+}$:Na$^+$/Li$^{+}$.  These materials have a low point-group symmetry, but the symmetry does guarantee that the principal axes of the spin Hamiltonian are aligned with the CF axes, so the Euler angles described in Sec.\ \ref{subsec:sh} are identically zero.  This avoids several technical challenges associated with C$_1$ point-group symmetry sites, while nevertheless demonstrating a crystal-field fit in a symmetry for which is is impossible to uniquely determine the crystal-field parameters from electronic energy (Stark level) data alone. The fit is performed using not only the Stark level data presented in Sec.\ \ref{subsec:laser_spect}, but also the ground-state Zeeman and hyperfine data detailed in Sec.\ \ref{subsec:epr}.

For $C_{2v}$ symmetry the crystal-field Hamiltonian, Eq.\ (\ref{eqn:c2v_cf_h}), contains eight non-zero parameters, which represent a distortion from cubic symmetry. During the fitting procedure, the eight parameters of Eq.\ (\ref{eqn:c2v_cf_h}) were varied, while the cubic $B_C^4$ and $B_C^6$ parameters were set to constant values that were previously determined for samples without Na$^+$ or Li$^+$ codopant \cite{wells_polarized_2000}. In addition to the $C_{2v}$ crystal-field parameters, the Slater parameters $F^k$ and the spin-orbit interaction parameter were varied, whereas the remaining parameters of Eq.\ (\ref{eqn:h_fi_defn}) were fixed to literature values. In order to accurately reproduce the ground-state hyperfine splittings, the coupling constant $a_l$ in Eq.\ (\ref{eqn:h_hf_defn}) was also varied. By diagonalizing the full Hamiltonian $H$ in the $\ket{\gamma L S J M_J I M_I}$ basis, and employing the projection defined by Eq.\ (\ref{eqn:operator_proj}), it was possible to determine the theoretical energy-level and spin Hamiltonian parameters for a given set of parameters.

In order to find a global solution for the parameters discussed above, the basin-hopping algorithm was used \cite{wales_global_1997,wales_global_1999}.  This procedure is similar to simulated annealing. However, after each step a conventional local minimization algorithm is employed.  Furthermore, unlike in simulated annealing, the ``temperature'' in the metropolis criterion is fixed to 1.  For the implementation employed here, the ``Bound Approximation by Quadratic Approximation'' algorithm from the NLopt library was used to perform the local minimization \cite{nlopt,powell2009bobyqa}.  The basin-hopping algorithm has the advantage of being less likely to become ``trapped'' in a local minimum compared to simulated annealing.  These calculations were completed with a program suite specifically developed for crystal-field analyses of low point-group symmetry hosts \cite{pycf}.

Using the procedure outlined above, a crystal-field fit was performed for the C$_{2v}$(Na$^+$/Li$^{+}$) centers in Sm$^{3+}$:CaF$_2$ and Sm$^{3+}$:SrF$_2$.  The results are split into three tables.  Table \ref{tab:eval} contains both the calculated energy level values as well as the experimental data obtained from site-selective excitation spectroscopy. Table \ref{tab:smcaf2srf2_shparam} summarizes the calculated and measured $g$-values and hyperfine parameters.
 
\clearpage

\begin{center}
  \begin{longtable}{cccccccccc}

    \caption{Experimental energy levels (cm$^{-1}$ in air, with uncertainty $\pm 1$ cm$^{-1}$) and theoretical energy levels (cm$^{-1}$) for the C$_{2v}$(Na$^+$/Li$^{+}$) site of Sm$^{3+}$ in alkaline-earth fluoride crystals. Experimental energies appended with a `?' denote uncertain assignments, whereas `-' denotes not observed Stark levels.} \\
   \label{tab:eval} \\
    \hline \hline
    Multiplet & State & \multicolumn{4}{c}{Sm$^{3+}$:CaF$_2$} & \multicolumn{4}{c}{Sm$^{3+}$:SrF$_2$} \\
    \cline{3-6} \cline{7-10}
    & & \multicolumn{2}{c}{Na$^+$} & \multicolumn{2}{c}{Li$^+$} & \multicolumn{2}{c}{Na$^+$} & \multicolumn{2}{c}{Li$^+$} \\
    \cline{3-4} \cline{5-6} \cline{7-8} \cline{9-10}
    & & Theory & Exp. & Theory & Exp. & Theory & Exp. & Theory & Exp. \\ 
    \hline
    \endfirsthead

    \caption[]{Experimental energy levels (cm$^{-1}$ in air, with uncertainty $\pm 1$ cm$^{-1}$) and theoretical energy levels (cm$^{-1}$) for the C$_{2v}$(Na$^+$/Li$^{+}$) site of Sm$^{3+}$ in alkaline-earth fluoride crystals. Experimental energies appended with a `?' denote uncertain assignments, whereas `-' denotes not observed Stark levels.} \\
    \hline \hline
    Multiplet & State & \multicolumn{4}{c}{Sm$^{3+}$:CaF$_2$} & \multicolumn{4}{c}{Sm$^{3+}$:SrF$_2$} \\
    \cline{3-6} \cline{7-10}
    & & \multicolumn{2}{c}{Na$^+$} & \multicolumn{2}{c}{Li$^+$} & \multicolumn{2}{c}{Na$^+$} & \multicolumn{2}{c}{Li$^+$} \\
    \cline{3-4} \cline{5-6} \cline{7-8} \cline{9-10}
    & & Theory & Exp. & Theory & Exp. & Theory & Exp. & Theory & Exp. \\ 
    \hline
    \endhead

    \multicolumn{10}{r}{Continued over page.} \\
    \hline \hline
    \endfoot

    \hline
    \endlastfoot
    
     $^{6}${H}$_{5/2}$   & Z$_1$    & 0     & 0     & 0     & 0     & 0     & 0     & 0     & 0     \\ 
                        & Z$_2$    & 32    & 41    & 51    & 60    & 45    & 50    & 59    & 72    \\ 
                        & Z$_3$    & 129   & 124   & 139   & 142   & 146   & -     & 164   & -     \\ 
     $^{6}${H}$_{7/2}$   & Y$_1$    & 940   & 956   & 949   & 972   & 971   & 967   & 973   & 993   \\ 
                        & Y$_2$    & 1148  & 1148  & 1145  & 1162  & 1121  & 1121  & 1132  & 1140  \\ 
                        & Y$_3$    & 1298  & 1319  & 1283  & 1304  & 1282  & 1269? & 1281  & 1283  \\ 
                        & Y$_4$    & 1350  & 1350  & 1371  & 1371  & 1310  & 1294? & 1336  & 1365  \\ 
     $^{6}${H}$_{9/2}$   & X$_1$    & 2179  & 2185  & 2185  & 2196  & 2201  & 2198  & 2198  & 2214  \\ 
                        & X$_2$    & 2339  & 2344  & 2331  & 2355  & 2313  & 2331  & 2333  & 2352  \\ 
                        & X$_3$    & 2357  & 2354  & 2354  & 2369  & 2356  & 2355  & 2355  & 2367  \\ 
                        & X$_4$    & 2518  & 2511  & 2514  & 2516  & 2492  & 2488  & 2495  & 2497  \\ 
                        & X$_5$    & 2546  & 2550  & 2559  & 2569  & 2513  & 2548  & 2535  & 2556  \\
     $^{6}${H}$_{11/2}$  & W$_1$    & 3511  & 3517  & 3506  & 3520  & 3519  & 3571  & 3519  & 3530  \\ 
                        & W$_2$    & 3550  & 3554  & 3555  & 3578  & 3572  & 3607  & 3581  & 3599  \\ 
                        & W$_3$    & 3743  & -     & 3731  & 3625  & 3710  & 3644  & 3734  & 3647  \\ 
                        & W$_4$    & 3754  & 3742  & 3760  & 3751  & 3739  & 3726  & 3737  & 3738  \\ 
                        & W$_5$    & 3781  & 3758  & 3777  & 3784  & 3763  & 3759  & 3759  & 3780  \\ 
                        & W$_6$    & 3900  & -     & 3901  & -     & 3860  & -     & 3879  & -     \\
     $^{6}${H}$_{13/2}$  & V$_1$    & 4850  & 4857  & 4852  & 4865  & 4871  & 4864  & 4879  & 4888  \\ 
                        & V$_2$    & 4893  & 4883  & 4902  & 4912  & 4926  & 4914  & 4933  & 4944  \\ 
                        & V$_3$    & 5004  & -     & 5010  & 5009  & 5002  & 5000  & 5020  & 5030  \\ 
                        & V$_4$    & 5203  & 5133? & 5187  & 5142  & 5162  & 5152  & 5161  & 5152  \\ 
                        & V$_5$    & 5209  & 5183? & 5201  & 5177  & 5181  & 5155  & 5177  & 5166  \\ 
                        & V$_6$    & 5233  & -     & 5229  & 5248  & 5208  & -     & 5224  & 5238  \\ 
                        & V$_7$    & 5268  & -     & 5253  & 5275  & 5245  & 5241  & 5252  & 5263  \\ 
      $^{6}${F}$_{1/2}$, & S$_1$    & 6165  & 6161  & 6167  & 6165  & 6189  & 6188  & 6190  & 6191  \\ 
      $^{6}${F}$_{3/2}$, & S$_2$    & 6193  & 6192  & 6209  & 6216  & 6217  & 6220  & 6228  & 6243  \\ 
      $^{6}${H}$_{15/2}$ & S$_3$    & 6286  & -     & 6294  & 6312  & 6302  & -     & 6335  & 6347  \\ 
                        & S$_4$    & 6572  & 6564  & 6575  & 6571  & 6540  & 6529  & 6550  & 6543  \\ 
                        & S$_5$    & 6703  & 6717  & 6693  & 6723  & 6672  & 6682  & 6659  & 6692  \\ 
                        & S$_6$    & 6737  & 6724  & 6723  & 6731  & 6686  & 6696  & 6690  & 6709  \\ 
                        & S$_7$    & 6743  & 6747  & 6735  & 6758  & 6710  & 6724  & 6725  & 6744  \\ 
                        & S$_8$    & 6760  & 6766  & 6762  & 6782  & 6727  & 6741  & 6744  & 6764  \\ 
                        & S$_9$    & 6847  & 6891  & 6832  & 6867  & 6807  & 6835  & 6802  & 6858  \\ 
                        & S$_{10}$ & 6901  & 6911  & 6882  & 6894  & 6850  & 6860  & 6855  & 6871  \\ 
                        & S$_{11}$ & 6927  & 6929  & 6921  & 6940  & 6877  & 6882  & 6894  & 6906  \\ 
      $^{6}${F}$_{5/2}$  & R$_1$    & 7253  & 7273  & 7259  & 7282  & 7229  & 7246  & 7239  & 7262  \\ 
                        & R$_2$    & 7377  & 7368  & 7376  & 7373  & 7335  & 7330  & 7338  & 7341  \\ 
                        & R$_3$    & 7396  & 7381  & 7404  & 7395  & 7356  & 7344  & 7367  & 7366  \\
      $^{6}${F}$_{7/2}$  & Q$_1$    & 8121  & 8130  & 8122  & 8136  & 8095  & 8108  & 8103  & 8120  \\ 
                        & Q$_2$    & 8130  & 8142  & 8130  & 8156  & 8107  & 8118  & 8115  & 8139  \\ 
                        & Q$_3$    & 8151  & 8162  & 8166  & 8173  & 8127  & 8135  & 8136  & 8154  \\ 
                        & Q$_4$    & 8275  & -     & 8275  & -     & 8239  & -     & 8249  & -     \\
      $^{6}${F}$_{9/2}$  & P$_1$    & 9284  & 9277  & 9283  & 9287  & 9258  & 9260  & 9266  & 9277  \\ 
                        & P$_2$    & 9292  & 9307  & 9296  & 9313  & 9269  & 9283  & 9278  & 9296  \\ 
                        & P$_3$    & 9316  & 9321  & 9319  & 9334  & 9291  & 9298  & 9305  & 9318  \\ 
                        & P$_4$    & 9375  & 9370  & 9378  & 9379  & 9340  & 9345  & 9357  & 9363  \\ 
                        & P$_5$    & 9386  & 9375  & 9388  & 9386  & 9359  & 9351  & 9371  & 9371  \\
      $^{4}${G}$_{5/2}$  & A$_1$    & 17657 & 17678 & 17672 & 17690 & 17704 & 17722 & 17716 & 17735 \\ 
                        & A$_2$    & 18074 & 18072 & 18067 & 18074 & 18070 & 18087 & 18046 & 18070 \\ 
                        & A$_3$    & 18101 & 18089 & 18107 & 18109 & 18121 & 18100 & 18146 & 18138 \\ 
                                                                                     
    \end{longtable}                                                                  
\end{center}

\begin{table}[tb!]
  \centering
  \caption{\label{tab:smcaf2srf2_shparam}%
  Fitted and experimental spin Hamiltonian parameters of the C$_{2v}$(Na$^+$/Li$^{+}$) site of Sm$^{3+}$ in alkaline-earth fluoride crystals. $A$-values are in cm$^{-1}$. It should be noted that due to the upper limit of the EPR magnetic field strength of 1.5 Tesla, the uncertainty of the hyperfine parameters was up to 80\%.}

  \begin{tabular}{ccccccccc}
    \hline \hline
    Parameter & \multicolumn{4}{c}{Sm$^{3+}$:CaF$_2$} & \multicolumn{4}{c}{Sm$^{3+}$:SrF$_2$} \\
   \cline{2-5} \cline{6-9}
    & \multicolumn{2}{c}{Na$^+$} & \multicolumn{2}{c}{Li$^+$} & \multicolumn{2}{c}{Na$^+$} & \multicolumn{2}{c}{Li$^+$} \\
    \cline{2-3} \cline{4-5} \cline{6-7} \cline{8-9}
    & Theory & Exp. & Theory & Exp. & Theory & Exp. & Theory & Exp. \\ 
    \hline

    $g_x$ & 0.608 & 0.606 & 0.584 & 0.583 & 0.609 & 0.606 & 0.635 & 0.637 \\
    $g_y$ & 0.301 & 0.300 & 0.300 & 0.300 & 0.346 & 0.340 & 0.335 & 0.330 \\
    $g_z$ & 0.476 & 0.477 & 0.558 & 0.558 & 0.537 & 0.537 & 0.560 & 0.557 \\
    $A_x$ & 0.025 & 0.021 & 0.026 & 0.022 & 0.022 &   -   & 0.025 & 0.024 \\
    $A_y$ & 0.011 & 0.020 & 0.008 & 0.020 & 0.014 &   -   & 0.012 &   -   \\
    $A_z$ & 0.008 & 0.007 & 0.010 & 0.010 & 0.011 &   -   & 0.012 & 0.010 \\
    \hline \hline

  \end{tabular}
\end{table}

\begin{table}[tb!]

  \centering
  \caption{\label{tab:smca2_srf2_params}%
  Fitted values for free-ion and crystal-field parameters of the C$_{2v}$(Na$^+$/Li$^{+}$) site of Sm$^{3+}$ in alkaline-earth fluoride crystals. For the case of Sm$^{3+}$:Na$^+$/Li$^+$:SrF$_2$, the hyperfine data was incomplete; consequently the coupling constant was fixed to the corresponding value obtained for the CaF$_2$ compounds. Parameters in square brackets were held constant during the optimization procedure.  The fixed $B_C^4$ and $B_C^6$ parameters correspond to values obtained by Wells and Reeves for the cubic centers \cite{wells_polarized_2000}.  Other parameters were fixed to the values obtained for Sm$^{3+}$:LaF$_3$ by Carnall \emph{et al}.\ \cite{carnall1989systematic}. 
}
  \begin{tabular}{ccccc}
    \hline \hline
    Parameter & \multicolumn{2}{c}{Sm$^{3+}$:CaF$_2$ (cm$^{-1}$)} & \multicolumn{2}{c}{Sm$^{3+}:$SrF$_2$ (cm$^{-1}$)} \\
    \cline{2-3} \cline{4-5}
    & Na$^+$ & Li$^+$ & Na$^+$ & Li$^+$ \\
    \hline

    $E_{\mathrm{AVG}}$ & 47486  & 47522  & 47488 & 47598 \\ 
    $F^2$              & 79010  & 79118  & 78625 & 79307 \\ 
    $F^4$              & 57172  & 57194  & 58490 & 57249 \\ 
    $F^6$              & 39566  & 39545  & 39027 & 39658 \\ 
    $\zeta$            & 1174   & 1172   & 1172  & 1173  \\ 
    $[B_C^4]$          & -2112  & -2112  & -1890 & -1890 \\
    $[B_C^6]$          & 945    & 945    & 776   & 776   \\
    $\Delta B^2_0$     & -127   & -142   & -193  & -231  \\
    $\Delta B^2_2$     & -9     & -34    & 0     & -40   \\
    $\Delta B^4_0$     & 38     & -3     & 158   & 372   \\
    $\Delta B^4_2$     & 132    & 225    & 169   & 272   \\
    $\Delta B^4_4$     & -43    & -66    & -121  & -121  \\
    $\Delta B^6_0$     & -340   & -439   & -208  & -59   \\
    $\Delta B^6_2$     & -3     & 2      & -34   & -88   \\
    $\Delta B^6_4$     & -228   & -199   & -200  & 58    \\
    $\Delta B^6_6$     & 203    & 450    & 0     & 138   \\
    $a_l$              & 0.0049 & 0.0046 &[0.0049]&[0.0046]\\
    \hline \hline

  \end{tabular}
\end{table}

As can be seen in Tables \ref{tab:eval} and \ref{tab:smcaf2srf2_shparam}, excellent agreement is found between both the energy level data and the $g$-value data.  The fit to the ground-state hyperfine values was not as accurate. However, it was well within the rather large experimental uncertainty for Sm$^{3+}$:Na$^+$/Li$^+$:CaF$_2$.  
In the fit, the $g$-value $\chi^2$ was weighted by a factor of $10^4$ compared to the Stark-level $\chi^2$, reflecting the difference in magnitude of the $g$-values relative to crystal-field splittings.
Due to the large experimental uncertainty for the hyperfine parameters,  
the $g$-value $\chi^2$ was weighted by a factor of ten compared to the $A$-value $\chi^2$ (for the hyperfine parameters in cm$^{-1}$). It was found that if the $A$-value contribution was increased beyond this, the predicted $g$-values differed considerably from the experimental values. The hyperfine data for the SrF$_2$ compounds is incomplete, so the hyperfine coupling parameter $a_l$ was fixed to the respective CaF$_2$ values.

\subsection{Analysis of crystal-field parameters}

Antipin \emph{et al.} \cite{antipin1972} used optical and EPR data to obtain crystal-field parameters for  orthorhombic Dy$^{3+}$ centres in fluorides. Their results are quite similar to ours.  
In particular, the fitted $B^2_0$ and $B^2_2$ crystal-field parameters are negative.
It was found that changing the sign of $B^2_0$ did not adversely affect the fit to Stark levels, but the resulting spin Hamiltonian parameters were inconsistent with experiment. For example, for Sm$^{3+}$:Na$^+$:CaF$_2$, changing the sign of  B$^2_0$, but leaving the other parameters as in Tab.\ \ref{tab:smca2_srf2_params}, gives a very different prediction for the spin-Hamiltonian parameters:  $g_x = 0.339$, $g_y = 0.778$, and $g_z = 0.404$, $A_x = 0.005$\,cm$^{-1}$, $A_y = 0.030$\,cm$^{-1}$, and $A_z = 0.015$\,cm$^{-1}$.  These observations mirror what was reported for Dy$^{3+}$ orthorhombic centres in fluorites. 

The intrinsic crystal-field parameters discussed in Section \ref{subsec:superposition} are always positive for (negatively charged) ligands \cite{New71,NN89a,NeNg00}, and Eq.\  (\ref{eq:sm}), along with the geometry discussed in Section \ref{subsec:superposition}, therefore gives $B_C^4 = -28/9 \bar{B}_4$ and $B_C^6 = \sqrt{3}\bar{B}_6$, in agreement with the signs obtained in our fit. 

In the  C$_{2v}$ sites a Ca$^{2+}$ ion on the $z$ axis is replaced by an Na$^+$ or Li$^+$, which, to a first approximation, is equivalent to a negatively charged ion at $|z|=2 R_0\sqrt{\frac{2}{3}}$ (Tab.\ \ref{tab:structure}).  In Cartesian coordinates we have
$C^2_0 = \sqrt{1/4}(3z^2-r^2)/r^2$, $C^2_{-2} = \sqrt{3/8}(x-iy)^2/r^2$, so from this simplistic analysis we
would expect a positive  $\Delta B_0^2$ and zero $\Delta B_2^2$.

Antipin \emph{et al.} \cite{antipin1972} argued that the negative value of $\Delta B^2_0$ and the non-zero  $\Delta B_2^2$ may be explained by a shift of the rare earth dopant towards the co-dopant alkaline-earth ion and an indirect exchange contribution mediated  the fluorine ligands.  In the context of the superposition model, a reduction of the intrinsic parameters for the two fluoride ligands closest to the alkali-earth would give a negative  $\Delta B_0^2$ and positive $\Delta B_2^2$. Changes in angles and distances due to distortions would further modify the parameters. In principle, the superposition model could  be used to estimate the distortion, as was done by Newman and Stedman for rare-earth ions in garnet hosts \cite{NS69}. However, in the case of the C$_{2v}$ centre discussed here, the number of degrees of freedom (of angles and ligand distances) made such an analysis intractable.  

We note that the deviations from cubic symmetry are  noticeably bigger for Li$^+$ co-doped materials. We attribute this to the larger ionic radius mismatch between the co-dopant and the lattice ions for Li$^+$ compared to Na$^+$. It is possible that the symmetry has been lowered by the off-axis location of the Li$^+$.

\section{Conclusion}

We have presented a spectroscopic study of the C$_{2v}$ centre in Sm$^{3+}$:Na$^+$/Li$^+$:CaF$_2$/SrF$_2$, including both site-selective excitation and fluorescence spectroscopy and EPR. The data was utilized to perform a detailed crystal-field analysis of the orthorhombic center, for all four materials. The addition of Zeeman and hyperfine data proved essential to giving an unambiguous set of crystal-field parameters, in particular obtaining accurate values for the $B^2_0$ parameters. Our results are largely consistent with earlier studies of the orthorhombic centers in Dy$^{3+}$ fluorides co-doped with alkaline earths. In agreement with that work, we confirm that the signs of the $B^2_0$ parameters are inconsistent with a simple electrostatic model. The analysis of this series of four materials, with a range of co-dopant and dopant ionic radius mismatches provides a unique dataset for exploring trends for a range of distortions of the cubic lattice symmetry. 


\subsection*{Acknowledgments}
 We would like to thank Dr.\ G.\ D.\ Jones for many useful discussions. We would also like to acknowledge Mr.\ W.\ Smith for technical support. SPH acknowledges financial support in the form of a Canterbury Scholarship by the University of Canterbury.

\section*{References}


\begin{thebibliography}{10}
\expandafter\ifx\csname url\endcsname\relax
  \def\url#1{{\tt #1}}\fi
\expandafter\ifx\csname urlprefix\endcsname\relax\def\urlprefix{URL }\fi
\providecommand{\eprint}[2][]{\url{#2}}

\bibitem{Guillot-NoelCalculationanalysishyperfine2010}
Guillot-No{\"e}l O, Le~Du Y, Beaudoux F, Antic-Fidancev E, Reid M~F, Marino R,
  Lejay J, Ferrier A and Goldner P 2010 {\em Journal of Luminescence\/} {\bf
  130} 1557--1565 ISSN 0022-2313

\bibitem{sukhanov2018}
Sukhanov A~A, Likerov R~F, Eremina R~M, Yatsyk I~V, Gavrilova T~P, Tarasov V~F,
  Zavartsev Y~D and Kutovoi S~A 2018 {\em Journal of Magnetic Resonance\/} {\bf
  295} 12--16

\bibitem{BakerEndor1970}
Baker J~M and Blake W~B~J 1970 {\em Proceedings of the Royal Society A\/} {\bf
  316} 63--80

\bibitem{antipin1972}
Antipin A, Davydova M, Eremin M, Luks R and Stolov A 1972 {\em Optika i
  Spektroskopiya\/} {\bf 33} 673--680

\bibitem{freeth_zeeman_1982}
Freeth C~A and Jones G~D 1982 {\em Journal of Physics C: Solid State Physics\/}
  {\bf 15} 6833 ISSN 0022-3719

\bibitem{falin2003}
Falin M~L, Gerasimov K~I, Latypov V~A, Leushin A~M, bill H and Lovy D 2003 {\em
  Journal of Luminescence\/} {\bf 102--103} 239--242

\bibitem{zhong_optically_2015}
Zhong M, Hedges M~P, Ahlefeldt R~L, Bartholomew J~G, Beavan S~E, Wittig S~M,
  Longdell J~J and Sellars M~J 2015 {\em Nature\/} {\bf 517} 177--180

\bibitem{usmani2010}
Usmani I, Afzelius M, de~Riedmatten H and Gisin N 2010 {\em Nature
  Communications\/} {\bf 1} 12

\bibitem{lovric2013}
Lovri{\'c} M, Suter D, Ferrier A and Goldner P 2013 {\em Physical Review
  Letters\/} {\bf 111} 020503

\bibitem{longdell_experimental_2004}
Longdell J~J and Sellars M~J 2004 {\em Physical Review A\/} {\bf 69} 032307

\bibitem{rippe2008}
Rippe L, Julsgaard B, Walther A, Ying Y and Kr{\"o}ll S 2008 {\em Physical
  Review A\/} {\bf 77} 022307

\bibitem{probst2013}
Probst S, Rotzinger H, W{\"u}nsch S, Jung P, Jerger M, Siegel M, Ustinov A~V
  and Bushev P~A 2013 {\em Physical Review Letters\/} {\bf 110} 157001

\bibitem{xavi2015}
Fernandez-Gonzalvo X, Chen Y~H, Yin C, Rogge S and Longdell J~J 2015 {\em
  Physical Review A\/} {\bf 92} 062313

\bibitem{weber_paramagnetic_1964}
Weber M~J and Bierig R~W 1964 {\em Physical Review\/} {\bf 134} A1492--A1503

\bibitem{ReevesSiteselectivelaserspectroscopy1992}
Reeves R~J, Jones G~D and Syme R~W~G 1992 {\em Physical Review B\/} {\bf 46}
  5939--5958

\bibitem{wells_polarized_2000}
Wells J~P~R and Reeves R~J 2000 {\em Physical Review B\/} {\bf 61} 13593--13608

\bibitem{MurdochSiteselectivespectroscopyTb1997}
Murdoch K~M, Jones G~D and Syme R~W~G 1997 {\em Physical Review B\/} {\bf 56}
  1254--1268

\bibitem{StricklandSiteselectivespectroscopyTm1997}
Strickland N~M and Jones G~D 1997 {\em Physical Review B\/} {\bf 56}
  10916--10929

\bibitem{zakharch.bp_experimental_1966}
Zakharchenya B and Rusanov I 1966 {\em Soviet Physics Solid State, USSR\/} {\bf
  8} 31--\& ISSN 0038-5654

\bibitem{jamison_sharp_1999}
Jamison S~P, Reeves R~J, Pavlichuk P~P and Jones G~D 1999 {\em Journal of
  Luminescence\/} {\bf 83–84} 429--434 ISSN 0022-2313

\bibitem{pack_ce3+:na+_1989}
Pack D~W, Manthey W~J and {McClure} D~S 1989 {\em Physical Review B\/} {\bf 40}
  9930--9944

\bibitem{yamaga_optical_2004}
Yamaga M, Yabashi S, Masui Y, Honda M, Takahashi H, Sakai M, Sarukura N, Wells
  J~P~R and Jones G~D 2004 {\em Journal of Luminescence\/} {\bf 108} 307--311
  ISSN 0022-2313

\bibitem{liu_electronic_2006}
Liu G 2006 Electronic {Energy} {Level} {Structure} {\em Spectroscopic
  {Properties} of {Rare} {Earths} in {Optical} {Materials}\/} ed Liu G and
  Jacquier B (Springer Science \& Business Media)

\bibitem{Reid2016}
Reid M~F 2016 {\em Handbook on the Physics and Chemistry of the Rare Earths\/}
  vol~37 ed B{\"u}nzli J~C and Percharsky V~K (North Holland) chap 284, pp
  47--64

\bibitem{carnall_systematic_1989}
Carnall W~T, Goodman G~L, Rajnak K and Rana R~S 1989 {\em Journal of Chemical
  Physics\/} {\bf 90} 3443--3457 ISSN 0021-9606, 1089-7690

\bibitem{wybourne_spectroscopic_1965}
Wybourne B~G 1965 {\em Spectroscopic properties of rare earths\/} (Interscience
  Publishers)

\bibitem{judd_operator_1963}
Judd B~R 1963 {\em Operator techniques in atomic spectroscopy\/} ({McGraw-Hill}
  New York)

\bibitem{mcleod_intensities_1997}
McLeod D~P and Reid M~F 1997 {\em Journal of Alloys and Compounds\/} {\bf 250}
  302--305

\bibitem{wells2004hyperfine}
Wells J~P~R, Jones G~D, Reid M~F, Popova M~N and Chukalina E~P 2004 {\em
  Molecular physics\/} {\bf 102} 1367--1376

\bibitem{New71}
Newman D~J 1971 {\em Advances in Physics\/} {\bf 20} 197--256

\bibitem{NN89a}
Newman D~J and Ng B 1989 {\em Reports on Progress in Physics\/} {\bf 52}
  699--763

\bibitem{NeNg00}
Newman D~J and Ng B~K~C (eds) 2000 {\em Crystal Field Handbook\/} (Cambridge:
  Cambridge University Press)

\bibitem{macfarlane_coherent_1987}
Macfarlane R~M and Shelby R~M 1987 {\em Spectroscopy of solids containing rare
  earth ions\/} ed Kaplyanskii A~A and Macfarlane R~M (North-Holland,
  Amsterdam) p~51

\bibitem{hurtubise_algebra_1993}
Hurtubise V and Freed K~F 1993 {\em Advances in Chemical Physics\/} {\bf 83}
  465--465

\bibitem{hurtubise_algebra_1993-1}
Hurtubise V and Freed K~F 1993 {\em Journal of Chemical Physics\/} {\bf 99}
  7946--7969 ISSN 0021-9606, 1089-7690

\bibitem{manthey_crystal_1973}
Manthey W~J 1973 {\em Physical Review B\/} {\bf 8} 4086--4098

\bibitem{wood_absorption_1962}
Wood D~L and Kaiser W 1962 {\em Physical Review\/} {\bf 126} 2079--2088

\bibitem{schoemaker_111-oriented_1966}
Schoemaker D 1966 {\em Physical Review\/} {\bf 149} 693--704

\bibitem{yamaga_role_2002}
Yamaga M, Honda M, Kawamata N, Samejima K and Wells J~P~R 2002 {\em {EPR} in
  the 21st Century\/} ed Kawamori A, Yamauchi J and Ohta H (Amsterdam: Elsevier
  Science {B.V.}) pp 201--206 ISBN 978-0-444-50973-4

\bibitem{wales_global_1997}
Wales D~J and Doye J~P~K 1997 {\em Journal of Physical Chemistry A\/} {\bf 101}
  5111--5116 ISSN 1089-5639

\bibitem{wales_global_1999}
Wales D~J and Scheraga H~A 1999 {\em Science\/} {\bf 285} 1368--1372 ISSN
  0036-8075, 1095-9203 {PMID:} 10464088

\bibitem{nlopt}
Johnson S~G The {NLopt} nonlinear-optimization package
  \url{http://ab-initio.mit.edu/nlopt}

\bibitem{powell2009bobyqa}
Powell M~J 2009 {\em Cambridge NA Report NA2009/06, University of Cambridge,
  Cambridge\/}

\bibitem{pycf}
Horvath S~P The pycf crystal-field theory package
  \url{https://bitbucket.org/sebastianhorvath/pycf}

\bibitem{carnall1989systematic}
Carnall W, Goodman G, Rajnak K and Rana R 1989 {\em Journal of Chemical
  Physics\/} {\bf 90} 3443--3457

\bibitem{NS69}
Newman D~J and Stedman G~E 1969 {\em Journal of Computational Physics\/} {\bf
  51} 3013--3023

\end{thebibliography}

\providecommand{\newblock}{}

\end{document}